\documentclass[twocolumn]{aastex63}
\usepackage{lineno}

\usepackage{svg}
\usepackage{amsmath}

\received{xx xx, 20xx}
\revised{\today}
\accepted{xx xx, 20xx}
\submitjournal{AJ}

\shorttitle{Radio Transient Detection with Closure Products and ML}
\shortauthors{Xia et al.}
\graphicspath{{./}{figures/}}
\begin{document}

\title{Radio Transient Detection with Closure Products and Machine Learning \footnote{Released on \today}}

\correspondingauthor{Xia Zhang}
\email{xia.zhang@research.uwa.edu.au }

\author[0000-0002-0786-7307]{Xia Zhang}
\affiliation{International Centre for Radio Astronomy \\  Research (ICRAR)\\
M468, UWA, 35 Stirling Highway, Crawley,  \\
Perth, WA 6009}

\author{Foivos I. Diakogiannis}
\affiliation{International Centre for Radio Astronomy \\ Research (ICRAR)\\
M468, UWA, 35 Stirling Highway, Crawley,  \\
Perth, WA 6009}
\affiliation{Data61, CSIRO,  Australia}

\author{Richard Dodson}
\affiliation{International Centre for Radio Astronomy \\ Research (ICRAR)\\
M468, UWA, 35 Stirling Highway, Crawley,  \\
Perth, WA 6009}

\author{Andreas Wicenec}
\affiliation{International Centre for Radio Astronomy \\ Research (ICRAR)\\
M468, UWA, 35 Stirling Highway, Crawley,  \\
Perth, WA 6009}



\begin{abstract}
\textcolor{black}{For transient sources with timescales of 1-100 seconds, standardized imaging for all observations at each time step become impossible as large modern interferometers produce significantly large data volumes in this observation time frame.}
\textcolor{black}{Here we propose a method based on machine learning and using interferometric closure products as input features to detect transient source candidates directly from the spatial frequency domain without imaging. We train a simple neural network classifier on a synthetic dataset of Noise/Transient/RFI events, which we construct to tackle the lack of labelled data. 
We also use the hyper-parameter dropout rate of the model to allow the model to approximate Bayesian inference, and select the optimal dropout rate to match the posterior prediction to the actual underlying probability distribution of the detected events.}
\textcolor{black}{The overall F1-score of the classifier on the simulated dataset is greater than 85\%, with the signal-to-noise at 7$\sigma$. The performance of the trained neural network with Monte Carlo dropout is evaluated on semi-real data, which includes a simulated transient source and real noise. This classifier accurately identifies the presence of transient signals in the detectable signal-to-noise levels (above 4$\sigma$)
with the optimal variance.}
\textcolor{black}{Our findings suggest that a feasible radio transient classifier can be built up with only simulated data for applying to the prediction of real observation, even in the absence of annotated real samples for the purpose of training.}

\end{abstract}


\section{Introduction} \label{sec:intro}
The temporal variation of radio transient sources contains a wealth of physical information about extreme dynamical events in the universe, normally on time scales from milliseconds to months or longer \citep{bhat2011searches}.  Effective and comprehensive wide time domain searches provide a powerful sample resource for the discovery of rare events. \textcolor{black}{The establishment of large radio interferometers provides unprecendented opportunities to detect new examples and classes of transient sources with their revolutionary combination of high sensitivity and wide Field of 
View \cite[hereafter  FoV; ][]{murphy2013vast}. However, the huge amount of data these telescopes collect poses a great challenge to computational and storage requirements \citep{SKA2020}. For example, the SKA low-frequency array (called SKA-low) generates correlation data at a rate of about 0.5 TB per second \citep{ska-parameter}. Processing these Exa-scale correlated signals by standard imaging would inevitably require extremely large computational resources. Creating images of the sky involves iterative operations of Fourier transformation and nonlinear deconvolution \citep{hogbom1974aperture,cornwell1999deconvolution}, which are complex and time-consuming tasks. To carry out the repeated imaging on timescales of 1-100s is beyond the planned SKA resources. 
It is thus impractical to image at each time step, or even close to this, for transient searching. 
Performing automatic and fast identification of signals with potential transients, and making images of only those triggers, is the only feasible solution to the above problem.}
\newline 

\textcolor{black}{The great challenge in generating triggers for transients is the ubiquitous presence of radio frequency interference (RFI). Different RFI events display different frequency and time characteristics. 
One type is persistent signals in a narrow frequency band. 
Another arises from broad band impulsive signals of artificial radio emissions in the near field of the radio telescope \citep{czech2018cnn}. This is called transient RFI and its time domain behaviour can mimic astronomical signals of transient. 
The occurrence of RFI far outweighs the occurrence of transient events.
A good detection pipeline therefore must reliably distinguish the transient RFI from astrophysical transients.}

\textcolor{black}{Some non imaging approaches for source detection (such as~\citealt{trott2011source};~\citealt{law2012all}) exist. These methods can avoid the time-consuming imaging procedure, which also inevitably create the imaging artefacts caused by the synthesized beam. 
\cite{law2012all} proposed a new technique based on the bispectrum, an interferometric closure quantity, to make the most of visibility information to discover variability in the sky. 
The bispectrum method is computationally simple and coherently sensitive to sources throughout the whole Field of View (FoV), thus has significant potential for wide FoV transient observations. 
In particular, since this method is calibration-independent, the data does not need to be calibrated before processing. 
Therefore closure products are a good candidate to act as an applicable `feature' for a transient detection pipeline.
\newline }

\textcolor{black}{Currently, artificial intelligence algorithms (AI) are increasingly being employed in radio astronomy and taking an important role to respond to the growing astronomical data volumes \citep{baron2019machine}. 
Progress has been made not only in the classification of radio morphologies \citep{wu2019radio}, but also in the identification of dynamic radio sky, which include radio frequency interference mitigation \citep{akeret2017radio, burd2018detecting}) and detection of various kinds of transient \citep{michilli2018single, connor2018applying, zhang2018fast, rowlinson2019identifying}. 
Many of these works used complex deep learning models to perform their prediction on the information of spectrograms (time versus frequency plots), and achieve very high performance, such as high accuracy. 
However, increasing the complexity of classification methods is not ideal for scenarios that require a simple and efficient approach. Moreover, uncertainty estimates are rarely provided in the current radio transient detection. The application of machine learning to the direct identification based on visibility information has yet to be explored.}
\newline 

\textcolor{black}{In this paper, we explore the application of machine learning to the direct identification of transients based on visibility information. More importantly, we focus on extending the improvement on performance (such as accuracy) to more informative predictions with uncertainty, which is inspired by a pioneering work from~\citet{gal2016dropout}. 
The author introduced using Monte Carlo (MC) dropout to perform Bayesian inference in a subset of models containing a distribution of neural networks, with posterior predictions. 
This approach has been used to inferring posteriors for predicting the visual galaxy morphology with uncertain labelling \citep{walmsley2020galaxy}. Our work here employs MC dropout in an effort to match the posterior predicted probability with the true underlying probability distribution of the events, which has not been used in model uncertainty optimization for radio astronomy.}

Thus, the following improvements are presented in our work: 
\begin{enumerate}
    \item We developed a deep learning pipeline where we use the closure product of the correlation data from radio arrays as the input to the network for detecting potential transient sources in the collected signal. This reduces the computational complexity of the problem.
    
    \item In order to address the problem of lack of  (human) annotated data with a large number of real transient examples,   we created a synthetic dataset corresponding to the features of the Australia Telescope Compact Array\footnote{The dataset will be open sourced upon publication of the manuscript}. 
    
    \item In order to get probabilistic estimates for the network predictions, we develop a method, with which we  match the model posterior probability to the underlying probability distribution of the detection events. That is, the output of the algorithm approximates the probability distribution of the event being a transient or the radio interference, or background noise.
    
    \item We verify the performance of the trained model, by testing it on new semi-real data that consists of real background sky observations and simulated transient events.
 
\end{enumerate}
Given the above, it is proven that supervised Deep Learning methods can be used with simulation data to detect transient signals even in the absence of annotated examples. The method presented here is cost-effective, realistic, and applicable for predicting potential transient events in a SKA transient source detection pipeline.

% In this chapter, we build up a deep learning classification network to detect whether there are signals of interest at all during the period of observation while maintaining the visibility information. The proposed combination of the bispectrum and the AI here allows us to deal with large data workloads while providing high detection accuracy yet with low false negative rate robustly. Once we make fast online classification, we can schedule the detailed follow-up analysis to the selected events. Thus the efficiency of whole pipeline could be improved. 

The rest of the manuscript is structured as follows: In Section~\ref{sec:Method}, we give detailed descriptions of the proposed machine learning strategy based on closure products, and also show the data set and the relevant labelling strategy that we used for this work. In Section~\ref{sec:Experiments and Results}, we evaluate the performance of our classifier trained on hard-labelled data set and investigate the model posterior probability matching to the underlying distribution of each event when trained on soft-labelled data set. The performance of the optimized model based on the same simulated data is presented in detail. Further semi-real data tests are conducted to demonstrate the detection capability of the optimized model. Conclusions are provided in Section~\ref{sec:Conclusion}.

\section{Methods} \label{sec:Method}
\subsection{Classification Task in Visibility} \label{sec:Strategy}

One of the specific difficulties in this task is to separate suspected transient signals from the noise-corrupted sky background signal and radio frequency interference (RFI) without the production of a sky-image. \textcolor{black}{Differential signal analysis, i.e. subtracting a reference from a target to remove the background, is a common strategy to study object variability in time domain astronomy \citep{riess1998observational, bramich2008new, offringa2014wsclean, stewart2016lofar, zackay2016proper}. 
Instead of subtracting images, we difference the visibilities.} The individual visibilities data dump, $V_{ij}(t_{k})$, is the output of correlation on the baseline between antennas $i$ and $j$. Hence the difference between the visibilities of two consecutive data dumps is: 
\begin{equation}
\label{eqn:Visibi-Diff}
\Delta V_{ij}(t_k) = V_{ij}(t_k)- V_{ij}(t_{k-1}).
\end{equation}
This approach leaves residual sources that were not present in the previous time-step, largely removing the relatively slow-changing sky background and thus improving the transients' detectability. 
%Therefore, over the observation time interval, the residual signals will appear as slow variabilities of existing sources contaminated with noise.
\newline

We use the closure product (also called bispectrum or closed triangle) as the desired feature of for machine learning, \citep{wilson2013interferometers}, which is formed by multiplying three correlated data streams from different baselines on a closed loop. We can write this as:
\begin{equation}
CP_{ijk}(t) = V_{ij}(t)\cdot  V_{jk}(t) \cdot V_{ik}(t)^{*} ,
\end{equation}
where $V_{ij}(t)$ follows the notation from the above, and $V_{ik}(t)^{*}$ is the conjugate of the output of correlation on the baseline between antennas $k$ and $i$. \textcolor{black}{This type of closure product is coherently sensitive to point sources anywhere in the FoV and is incoherent for local interference \citep{law2012all, cornwell1987radio, kulkarni1989self, rogers1995fringe}. Thus in closure products, transients and RFI behave in different ways. The transient source is a delta function in the sky with no dependence on closure triangles, which would be the same for all closure products. RFI resolves on longer baselines as the source is located in the near-field of the array. Thus the signals do not have a planar wavefront, but are expanding ripples and the correlated signal degrading with antenna separation. The bispectrum is also calibration independent. It can be applied to improve the quality of visibility functions by canceling all instrumental residual phase terms, such as receiver delays and phases \citep{wilson2013interferometers}, thus making it computationally efficient for the detection of transients in raw data from the correlator.}  
\newline

The major problem with closure products is that they are sensitive to signals only with significant SNR in the bispectrum per closure product. This is because the closure product is formed by the multiplication of several visibilities, which implies that we might lose some weaker transients. But in fact, it is impractical to attempt to observe such weak events in all cases. If the signals are too weak, it means that we need more collection time, which is not possible for transients and the signal would not reach detectable levels for any method. Therefore, we can expect only the non-faint pulses as detectable targets (see discussion in \citealt{dodson2014kava}). This ensures that the closure product is the optimal responsive strategy to be applied in transient detection with radio interferometers. 
\newline

\begin{figure*}
\gridline{\fig{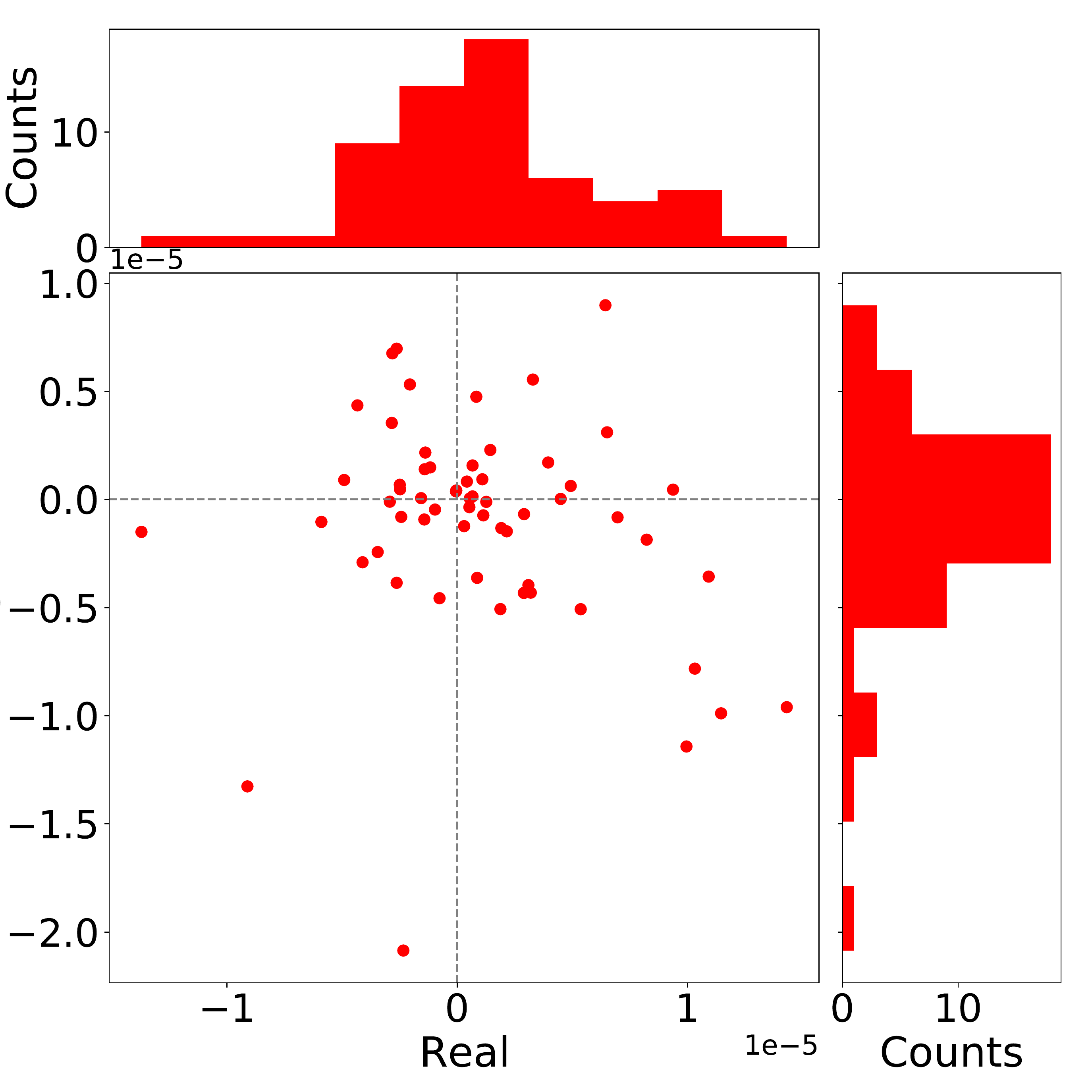}{0.33\textwidth}{\label{fig:Simulated noise} (a) Simulated noise. Red dots represent the closure products of noise.}
          \fig{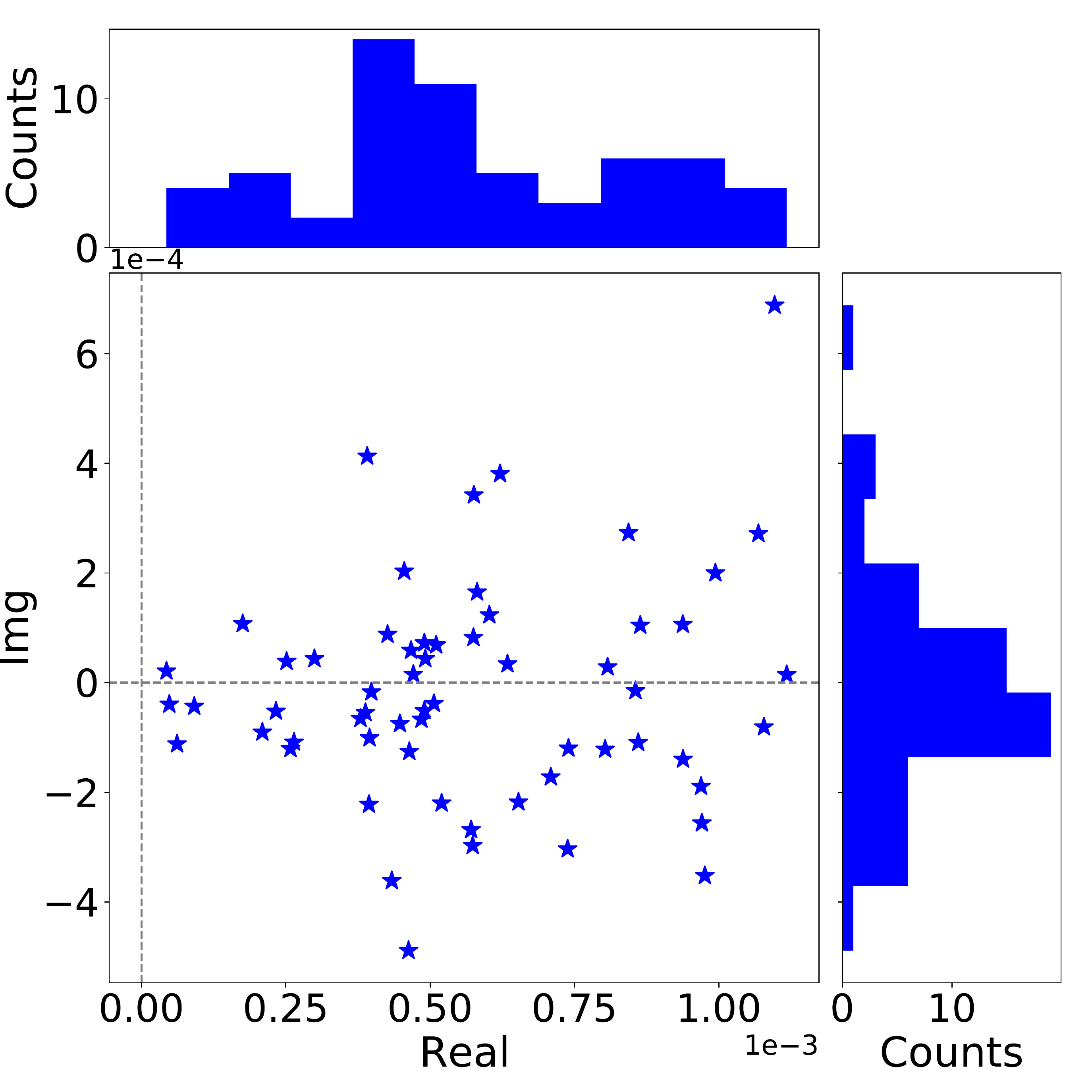}{0.33\textwidth}{\label{fig:Simulated transient}(b) Simulated transient. Blue stars represent the closure products of  transient.}
          \fig{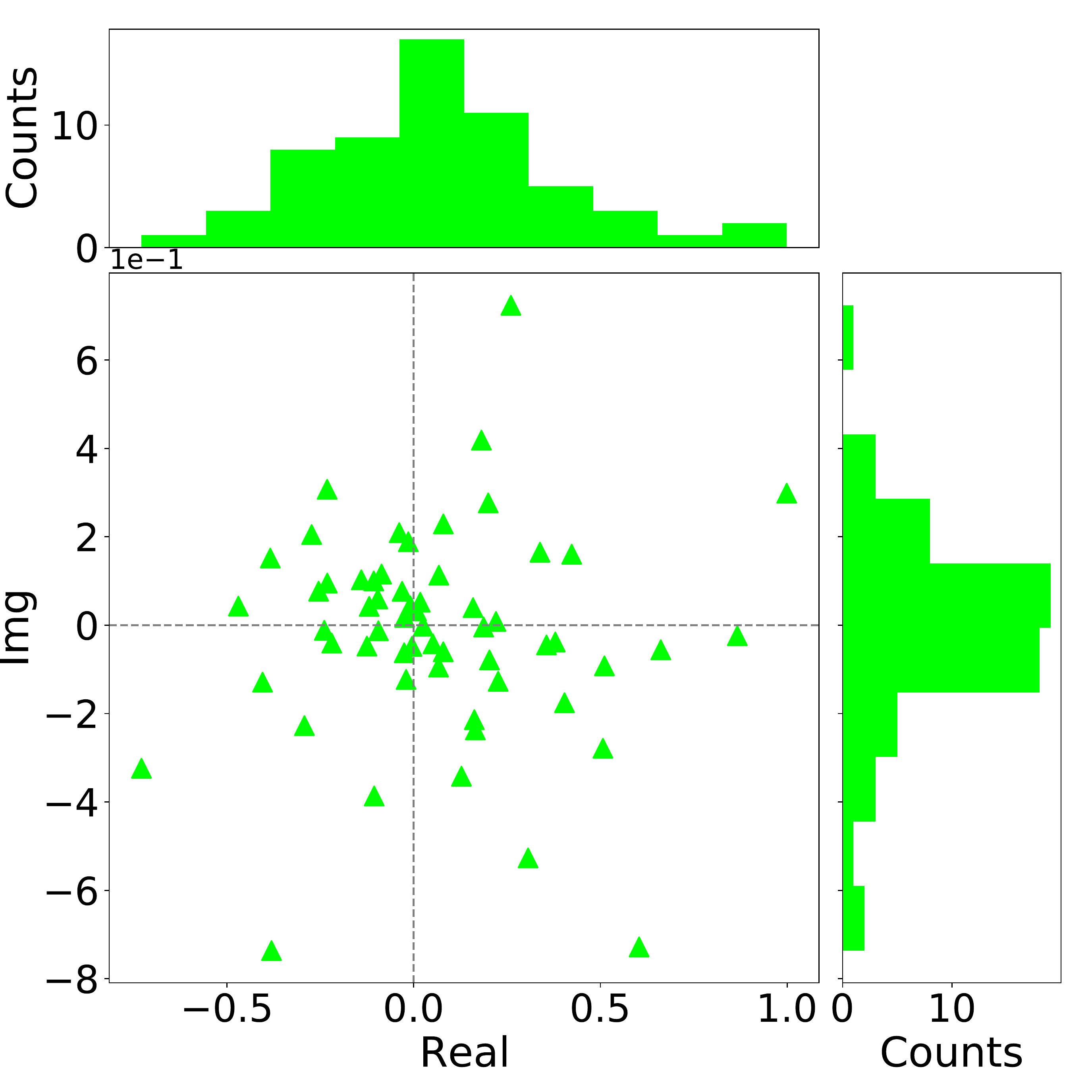}{0.33\textwidth}{\label{fig:Simulated RFI}(c) Simulated RFI. Green triangles represent the closure products of the RFI.}
          }
\caption{Examples of the distributions of the time differencing of closure products of the three components in the complex plane at a SNR level of $7\sigma$.}.
\label{fig:ThreeComponents}
\end{figure*}

\textcolor{black}{The residual of the closure product in our data files could be defined as:}
\begin{equation}
\label{eqn:CP-Diff}
\Delta CP_{ijk}(t) \equiv  \Delta V_{ij}(t)\cdot \Delta V_{jk}(t)\cdot \Delta V_{ik}(t)^{*}.
\end{equation}
The characteristics of the bispectrum residual of noise, RFIs and transients in data segments are listed below. The real and imaginary components representation of the complex information in Figure~\ref{fig:ThreeComponents} show the distinction among these three main catergories of components.

\begin{itemize} 
  \item{Noise}: the complex of bispectrum reside near the origin of the real-imaginary plane.
 
   \item{RFI}: the bispectrum of RFI shifts (potentially) to an arbitrary value and is widely spread across the complex plane, since the emissions of RFI are near-field or multi-path propagations,  which generate random complex information.
 
\item{Transients}: the bispectrum of transient migrate toward positive real values with a small spread. This is true for all of the closure products, while such behaviour is not the case for RFI. 
\end{itemize}

\subsection{Artificial Neural Network}
To detecting transients we trained a supervised machine learning model to classify the signal components from the data stream automatically and efficiently. Neural networks are a class of very flexible and non-linear models to perform classification tasks. 
The Artificial Neural Network (hereafter ANN) we use simply consists of an input layer, multiple hidden layers, and an output layer, which are all fully connected. In the prediction process of the model, when the input signal is received at the input layer, the information flows in the direction of the arrow in the network model diagram to the output layer unidirectionally, a process called “forward pass”. The outputs of the output layers are compared with their true labels to calculate a defined distance, bias or loss.

\textcolor{black}{Our MLP architecture is listed in Table~\ref{table:MLParchitecture}. To simplify our classification model as much as possible, we have only three hidden layers in our MLP model, and they are fully connected to the input and output layers. The ReLU is used as the activation function after each hidden layer;  the Softmax is used as the activation function for the output layer.} \textcolor{black}{We use Pytorch to implement the MLP and accelerate the training of the MLP via NVIDIA RTX 3090 GPU. We used the Adam optimizer \citep{kingma2014adam} for training, which is an extension of the stochastic gradient descent method. The initial learning rate is set to $\alpha = 10^{-3}$ and the batch size is set to 64. The MLP is then trained on the training set, and the training period is set to 1000 epochs when the model has reached convergence.}
\newline

\begin{table}[h!]
\centering
\begin{tabular}{|l|l|l|}
\hline
Layer (type) & Output shape    & Parameters \\ \hline
Linear-1     & {[}-1, 1024{]} & 123904    \\ \hline
Dropout-1    &                 & 0          \\ \hline
ReLU       &                 & 0          \\ \hline
Linear-2     & {[}-1, 2048{]} & 2099200  \\ \hline
Dropout-2    &                 & 0          \\ \hline
ReLU        &                 & 0          \\ \hline
Linear-3     & {[}-1, 256{]}   & 524544    \\ \hline
Dropout-3    &                 & 0          \\ \hline
ReLU        &                 & 0          \\ \hline
Linear-4     & {[}-1, 3{]}   &  771    \\ \hline
Softmax       &                 & 0          \\ \hline
\end{tabular}
\caption{\textcolor{black}{ The MLP architecture. ``Linear" refers to a fully connected layer. The ReLU is the activation function after each hidden layer; and the Softmax is the activation function for the output layer.} Total parameters: 2748419; trainable parameters: 2748419; nontrainable parameters: 0.}
\label{table:MLParchitecture}
\end{table}

\subsection{Monte Carlo Dropout to Approximate Bayesian Neural Network}

Unlike deterministic prediction with fixed weights, the Bayesian methods treat the model parameters as random variables when making predictions. Given a training dataset $\mathbb{D}$, a model distribution can be built that contains multiple weights with different parameters. This distribution is referred to in the Bayesian framework \citep{jaynes2003probability} as the posterior distribution, denoted as $p(\omega|\mathbb{D})$, where $\omega$ is a set of weight parameters of the model. 
\newline

To present the prediction uncertainty of the model more clearly, we consider the distribution of the probabilities of the softmax output of category instead of their mere pseudoprobability values. The variance on the predicted category probabilities should be more informative than presenting the uncertainty of the integer  labels. Let a new input be $x^{*}$, and the corresponding softmax layer output probability vector be $y^{*}$, which contains three prediction category probabilities, being given as
\begin{equation}
    p(y^{*}| x^{*}, \mathbb{D}) = \int p(y^{*}| x^{*}, \omega )\ p(\omega |\mathbb{D})\ \textup{d}\omega.
\end{equation}

Logical analysis of the weight posterior distribution $p(\omega|\mathbb{D})$ is usually infeasible, especially for complex structures like neural networks. However,~\cite{gal2016dropout} applies `dropouts', a regularization technique, to analyze the uncertainty of a model and shows that it can be approximated as a Bayesian neural network.
The dropout rate $p$ is a hyperparameter of the model, taking a value between 0 and 1, which reflects the probability of each neuron being dropped out during the implementation. The stochastic behavior of each neuron follows the same Bernoulli distribution. As mentioned above, neuron rejection generally occurs during the training phase. Here, the dropout is also turned on in the prediction phase. Performing multiple prediction passes completes the Monte Carlo sampling in the approximate distribution of $p(\omega|\mathbb{D})$, thus
\begin{equation}
    \begin{split}
    p(y^{*}| x^{*}, \mathbb{D}) 
    & = \int p(y^{*}| x^{*}, \omega )p(\omega |\mathbb{D})\ \textup{d}\omega   \\
    & \approx \frac{1}{T} \sum_{t = 1}^{T}p(y^{*}|x^{*},\omega _{t}),
    \end{split}
\end{equation}
where $T$ is the number of stochastic forward passes through the neural network using dropout.
The above method is known as the Monte Carlo Dropout approximation. More specific details can be found in ~\cite{gal2016dropout}. 
\newline 

With the Monte Carlo Dropout, we can analyze the uncertainty of our model. By comparing uncertainties of models with different dropout, we can select the appropriate value of the dropout rate. In many papers on practical applications 
\citep[e.g. ][]{walmsley2020galaxy}, 
the dropout value is chosen as 0.5 or 0.25 for convenience, and such a common choice does reduce the overfitting of the model. A value like 0.5 can give good classification results on more complex networks. However, networks without calibration containing dropout are not optimized. In particular, for networks with different applications, their structures and depths are not quite the same. If the same dropout rate is used \textcolor{black}{for matching the uncertainty of the predictions} without optimization, the optimal performance of the classifier cannot be guaranteed. Our investigation, in addition to finding the optimal dropout rate to maximize the score for detection, also provides probability matching.

\subsection{Data set descriptions}

\textcolor{black}{Since the number of real transients on the aforementioned time scale is relatively low, we lacked labelled datasets containing a large number of short-duration transients to train our model,} which means annotated data did not exist and thus we used simulations. We created a range of simulated datasets for use with our model. The one is purely simulated data, which was used to train the transient classifier and to investigate the feasibility of the dropout-base Bayesian network. The other was semi-real data, which was a real MWA observation that provided a background sky to which we added simulated transients to verify the performance of the well-trained neural network.
\newline 

The telescope configuration used for the pure simulation matched the Australia Telescope Compact Array (ATCA). The ATCA has sufficient baselines to develop the model in a rapid manner. 
It would be trivial, but time consuming, to scale this work up to the SKA but there are no obvious benefits from this. We simulated observations of the six antennas of ATCA. 
\textcolor{black}{In our simulations, we observed a local sky area of RA=01h11m13.0s and DEC=-45d45m00.0s using the central frequency of 1.42\,GHz and a field of view of 15 arcmins wide. 
100 different static sky models were produced in 10 surrounding frequency channels with a width of 0.1 MHz.} Each sky model contains a random number of 4 to 8 sky background sources with random locations and varying strength.
We randomly selected one of the 100 sky models and added a point source at a random location.
The parameters of the transients include arrival time of the pulse and a random width (between 1 and 10-time steps). 
The RFI was simulated with a random (complex) number added to the baseline data at a random time step. 
The correlated data streams produced were then differenced, and then the closure product formed with the residual information (see Equation~\ref{eqn:CP-Diff}). 

The noise in practice here contains two aspects. One is the ``thermal noise" of the observational system. The other is the ``measurement error", which comes from the simple differencing method we implemented.
``Measurement error" could be reduced by interpolating over points, rather than differencing. Although an interpolation method would be more accurate compared to the differencing method, in practice, the errors introduced by both are swamped by the ``thermal noise".
We keep the ``measurement error" and the ``thermal noise" at a constant level and gradually suppress the signal-to-noise ratio to test the ability of the machine learning model. Our simulated ``thermal noise" is generated by a Gaussian distribution with a standard deviation of 0.01 arbitrary units in strength, and the measurement error is kept at the same level of intensity as the thermal noise.
\newline

The high SNR transient is well characterized and easy to find, but the weak signals will pose a challenge to model training and prediction. In our experiments, we trained and tested several datasets with different SNR quality. Here we present only the dataset with the lowest SNR that maintains the high performance of the model, which we call the high SNR dataset; and the dataset with the lowest SNR that has significantly decreased the performance of the model, yet still can be a good training dataset to maintain a working model, which we call the low SNR dataset. For the high SNR dataset, the SNR ratio is 7$\sigma$, which includes a Gaussian distribution simulation with a transient source intensity of 0.1, ``thermal noise" and ``measurement error" of 0.01. For the low SNR dataset, the SNR ratio is 4$\sigma$, where the noise is kept constant, and the transient source intensity is reduced to 0.06.
\newline

\textcolor{black}{In each observational data segment, time steps containing transient sources, RFI, and noise are extracted respectively to form an available group of simulated samples. For each of these sample data, we selected three frequency channels with 20 different closure products in each frequency channel. Since the machine learning model deals with real numbers, we separate the real and imaginary parts of the data and then ``deconvolve" the 3 $\times$ 20 complex data points of each sample into a real vector of length 120.} We set the closure product $CP_{1f} = a_{1f} + i b_{1f}$ in one frequency channel with 20 closure product components.
Thus the input vector is:
\begin{equation}
(a_{1f}, b_{1f}, a_{2f}, b_{2f},a_{3f},b_{3f}).
\end{equation}

We generated a total of 5,000 such sets of simulated samples for each category. They are labeled and then fused into a large dataset containing a total of 15,000 data samples.
Three splits of the data into training, validation and test set, are required to train and evaluate the MLP classification.
One-tenth of the data (which contains equal amounts of the three signal components of transient source, RFI, and noise) is taken as the validation set, another tenth is taken as the test set, and the remaining datasets were used as the training set. 
\newline

% \begin{figure}
% \gridline{\fig{a_sample_of_closureProduct_of_noise.pdf}{0.33\textwidth}{\label{fig:Simulated noise}Simulated noise. Red dots represent the closure products of noise.}
%           \fig{a_sample_of_closureProduct_of_transient.pdf}{0.33\textwidth}{\label{fig:Simulated transient}Simulated transient. Blue stars represent the closure products of  transient.}
%           \fig{a_sample_of_closureProduct_of_RFI.pdf}{0.33\textwidth}{\label{fig:Simulated RFI}Simulated RFI. Green triangles represent the closure products of the RFI.}
%           }
% \caption{Examples of the distributions of the time differencing of closure products of the three components in the complex plane at a SNR level of $5\sigma$.}.
% \label{fig:ThreeComponents}
% \end{figure}

\subsection{Labelling of Samples}
In the first experiment, we will build up the simple ANN to perform a simple classification.
Following the usual labeling convention, we label 0, 1, and 2 of  three categories as noise, transients, and RFI, respectively. These are the hard integer labels for the samples.
\newline

In the second experiment, we improved the representation of categorical data in prediction analysis by changing the way to encode categorical variables. The predicted output of the model after the softmax activation layer is retained as a probability vector instead of the one-hot label. Meanwhile, we also use the probability vector soft labels to label the samples of the data. 
As implemented, we attribute a certain degree of uncertainty to the probability of each category in the vector after converting the one-hot binary labels into probability label vectors, which approximates the uncertainty of the data and facilitates the comparison between the predicted output of the probability vector and the true labels of the probability vector.
Such manipulation allows the representation of categorical data to be more expressive and provides more detailed information than a single hard label.
\newline

\textcolor{black}{
For each training epoch, the probability values are randomly drawn from some probability distribution, which is the ground truth distribution of the soft labels. In practice, in each training epoch, for each instance, after converting its hard label into an one-hot label vector, the element at the position corresponding to its own category is a probability randomly extracted from the normal distribution (2/3, 0.01) as the value. In contrast, two probability values are assigned to the corresponding heterogeneous category from the normal distribution (1/6, 0.01) at the position of the two classes. Then the formed probability vector is normalized and corrected to satisfy the sum of 1 so that each instance is given a suitable soft label. Thus, the experimental procedure is to train the model from data with random probability soft labels. We propose to determine the optimal dropout rate by analyzing the statistical approximation of the probabilistic output distribution of each predicted data about the category to its original true probabilistic label distribution. }
\newline

\subsection{Evaluation Metrics for Posterior Distribution}
\textcolor{black}{
The probability output of the classifier for each instance includes three classes of probability distributions. To evaluate the performance of these probability outputs, we can examine the similarity of the output probability distributions to the input ground truth distributions in two different ways. The first is that we generalize the three categories of probability distributions by overlapping them into a one-dimensional curve of the output distribution. In this way, we can then use the commonly used KL-divergence that measures the distance between two 1-dimensional distributions. KL-divergence is defined as follows:
\begin{equation}
    D_{KL}(q(\widehat{s})|| p(s)) = \sum_{t}^{T} q(\widehat{s_{t}})\ \textup{log}\frac{q(\widehat{s_{t}})}{p(s_{t})},
\end{equation}
where $s$ is the soft label of ground truth and $\widehat{s_{t}}$ is the soft label of prediction.}
\newline

\begin{figure}[]
\centering
\includegraphics[width=0.49\textwidth]{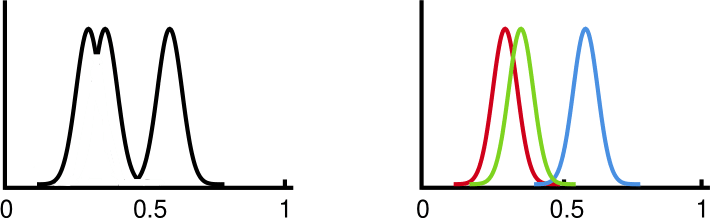}
\caption{\label{fig:1d-3d}  Posterior prediction with 1-d overall distribution vs. 3-d distributions.}
\end{figure}

\textcolor{black}{
However, applying only KL-divergence to analyze the overall distribution cannot take into account the origins of those sub-distributions. When the multiple distributions in the right panel of Figure~\ref{fig:1d-3d} are superimposed into one distribution, the KL-divergence cannot distinguish between color class distributions (the left panel in Figure~\ref{fig:1d-3d}). This would result in the KL-divergence making the same judgments,  even if we switch the labels corresponding to the color distributions (i.e., it would no longer match ground truth). Therefore, we added an evaluation metric to ensure a trustworthy analysis.}
\newline 

% \begin{figure}[]
% \centering
% \includegraphics[width=0.49\textwidth, trim={1.5cm 1.8cm 3cm 0.5cm}]{figures/KL-dice.png}
% \caption{\label{fig:1d-3d} \foivos{Make it single column figure (smaller).} Posterior prediction with 1-d overall distribution vs. 3-d distributions.}
% \end{figure}

\textcolor{black}{
The second way is to keep the soft label output as a multivariate probability distribution and use the Dice distance \citep{dice1945measures} to measure the dissimilarity of the distribution across multiple dimensions (see representations in the right panel of Figure~\ref{fig:1d-3d}. Dice distance is a simple and useful aggregated measure of spatial overlap in a multi-category context in semantic segmentation. It is complementary to the Dice similarity coefficient (DSC) and is obtained by subtracting the DSC from 1 \citep{dice1945measures}. There are multiple definitions for Dice coefficient \citep{milletari2016v, sudre2017generalised}. Since they are numerically equivalent, we choose one of the definitions, which is defined as follows:
\begin{equation}
D(\mathbf{p},\mathbf{l}) = \frac{\sum_{i}^{}p_{i}l_{i}}{\sum_{i}^{}(p_{i}^{2} + l_{i}^{2}) -\sum_{i}^{}p_{i}l_{i}},
\end{equation}
where $\mathbf{p} \equiv {p_{i}}$ and $p_{i} \in \left [0,1 \right]$ is a set of continuous-valued probabilistic outputs. And therefore the Dice distance is $1-D(\mathbf{p},\mathbf{l})$. In semantic segmentation, $p_{i}$ represents the probability vector of the $i$-th pixel. Here we can let it represent the set of predicted probability distributions for the location of the $i$-th element in the soft label, and $\mathbf{l} \equiv {l_{i}}$ is the corresponding set of ground truth label distributions. The Dice distance is minimum when $\mathbf{p}\rightarrow \mathbf{l}$.}
\newline

\section{Experiments and Results} 
\label{sec:Experiments and Results}
In this section, we perform an experimental analysis of the inference performance of the transient source detection pipeline constructed with hard labels and the dropout-based neural network constructed with soft labels.

\subsection{The classification using ANN with Hard Labels}

\begin{figure*}[!htb]
\gridline{\fig{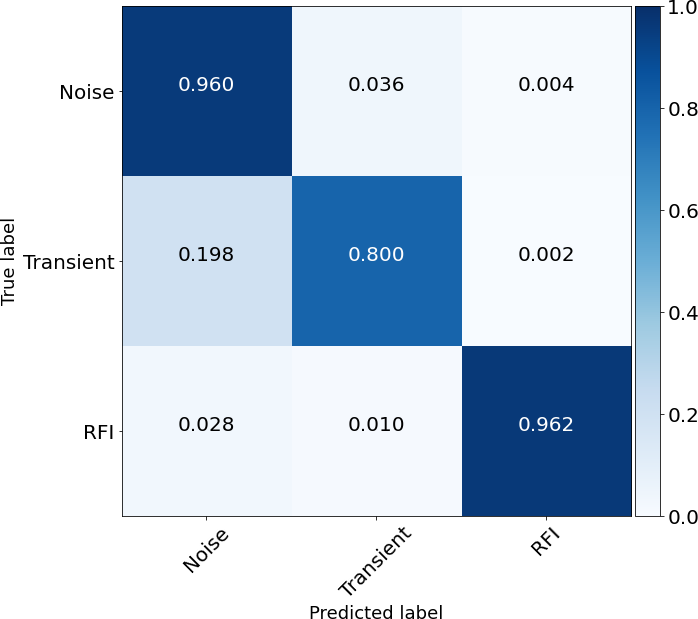}{0.48\textwidth}{\label{subfig:high high}(a) High train + high test}{}
          \fig{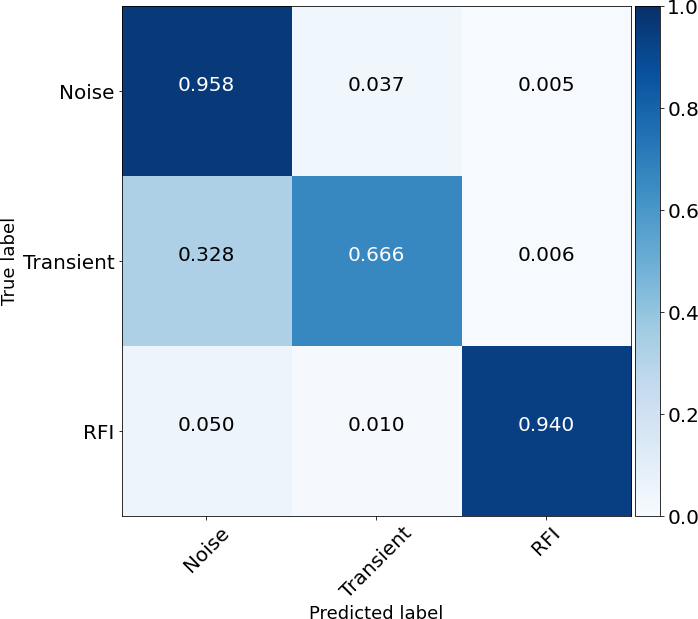}{0.48\textwidth}{\label{subfig:high low}(b) High train + low test}{}
          }
          
\gridline{\fig{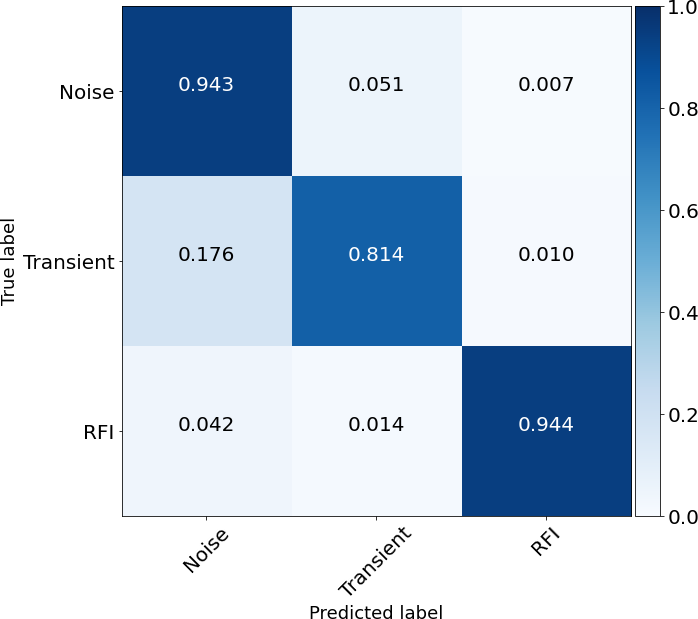}{0.48\textwidth}{\label{subfig:low high}(c) Low train + high test}{}
          \fig{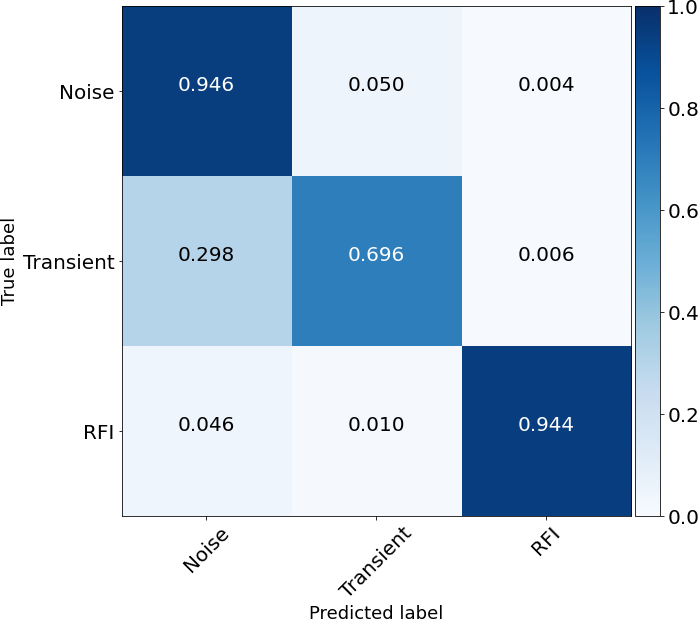}{0.48\textwidth}{\label{subfig:low low}(d) Low train + low test}{}
          }
\caption{Confusion matrices for the three classes on simulated test datasets with different SNR (low SNR level of $3\sigma$ and high SNR level of $7\sigma$). Along each row, the MLP classifiers are trained with the training data set at the same SNR; along each column, the well-trained classifiers are tested with the test data at the same SNR.}
\label{fig:Cross high and low snr}
\end{figure*}

The results of the neural network on the  high SNR level (7$\sigma$) we reached can be seen in Figure~\ref{fig:Cross high and low snr}(a), where the columns represent the instances of the predicted classes, and the rows represent the instances of the actual classes. The diagonal cells in these matrices contain percentage values, indicating the proportion of samples correctly assigned, while the non-diagonal cells contain error rate values. The model can do an excellent job in identifying noise and RFI, as they yield classification scores higher than 96\%. The classification performance can be considered a promising result despite the confusing inference in distinguishing transients from noise. At such a low SNR level, our transient signal is classified as noise 19.8\% of the time. On the other hand, if it is noise, there is only a 3.6\% probability of being classified as transient and not likely to be confused with RFI. It clearly shows that our idea is effective for classifying RFI, transients, and noise.
\newline

\textcolor{black}{Since the quality of the real data in the task to be classified is hardly the same as the one of the training data used for training the classifier, we used both high (7$\sigma$) and low (4$\sigma$) SNR training data to train the classifiers separately, and then cross-test the low and the high SNR validation datasets to explore the best model training and validating process.} Specifically, for example, in Figure~\ref{fig:Cross high and low snr}(b), during training with a high SNR (7$\sigma$) dataset, we used \textcolor{black}{12,000 samples for training and 1,500} ones for validating. The parameters of the model were saved according to the best MCC overall score, which includes all the entries (namely false positives, false negatives, true positives, and true negatives) of the confusion matrix in its calculation, and is generally regarded as a reliable, balanced measurement \citep{grandini2020metrics}. Then the inference was performed with a low SNR test dataset (including 1,500 samples). For Figure~\ref{fig:Cross high and low snr}(c), the operation was just the opposite. And for the experiments on the diagonal, we performed regular training and testing with the same SNR dataset.
\newline

With the high SNR test data, we found that the model trained at low SNR (4$\sigma$) predicted transients with slightly higher accuracy than the model trained at high SNR (7$\sigma$) dataset (comparison between the centers of panels).  While for the prediction of noise and RFI, the model trained at low SNR (4$\sigma$) is worse than the model trained at high SNR (7$\sigma$). Most of the reduction in noise prediction (96\% to 94\%) and some of the misclassification of RFI go into the confusion of transients, which causes an increase in false positives for transient prediction (comparison along the second column in Figure~\ref{fig:Cross high and low snr}(a), \ref{fig:Cross high and low snr}(c)). Similar false detections of noise as transients are also more likely to occur when the SNR of the test data decreases (second column comparison in Figure~\ref{fig:Cross high and low snr}(b) and \ref{fig:Cross high and low snr}(d). Although the prediction accuracy of transients is a metric of interest, we prefer to minimize the occurrence of noise and RFI false detection rates as transients. Therefore, we favour the model under high signal-to-noise data training.
\newline 

Furthermore, comparing along the rows in Figure~\ref{fig:Cross high and low snr}(a) \& \ref{fig:Cross high and low snr}(b), and Figure~\ref{fig:Cross high and low snr}(c) \& \ref{fig:Cross high and low snr}(d), we find that the model trained with high SNR (7$\sigma$) also controls false positives well when applied to different SNR test datasets. In contrast, the model trained at low SNR (4$\sigma$) yields surprisingly higher false positive rates as the SNR of the test data increases. In order to tackle the slightly higher false negatives caused by noise and the slightly lower true positives for transients, we will suggest ways to improve the SNR of the observations operationally. Furthermore, a method of evaluating uncertainty in the model inference can also help us to exclude some unreliable predictions, as will be discussed next.

\subsection{Bayesian Experiments}

In this experiment, the batch size, initial learning rate, and stochastic gradient descent optimization method among the training retained the training settings of the original experiments, and the training process lasts for 1,000 epochs. After training, the training result of the iteration with the best global metric MCC performance is saved as the best model. We applied dropout to the prediction process by implementing 100 Monte Carlo forward passes, which meant we had 100 Monte Carlo approximation networks that produced 100 posterior responses to each categorical element in predicted  probability vector for each test data.
\newline 

\textcolor{black}{
Here for the dropout rate to be evaluated, we choose the random search method, which is a common method of hyper-parameter tuning for machine learning. When multiple hyper-parameters to be tuned in machine learning, there are tens of thousands of combinations required. Random search is a more efficient and fast strategy than the traditional grid search method \citep{bergstra2012random}. We also apply this strategy to our study of finding the optimal dropout rate. Although it is a one-dimensional search, as shown in the result, it is the application of the random search strategy that allows us to find the optimal dropout rate in the logarithmic space of parameters more quickly.}
\newline

\begin{figure}[h]
\centering
\gridline{\fig{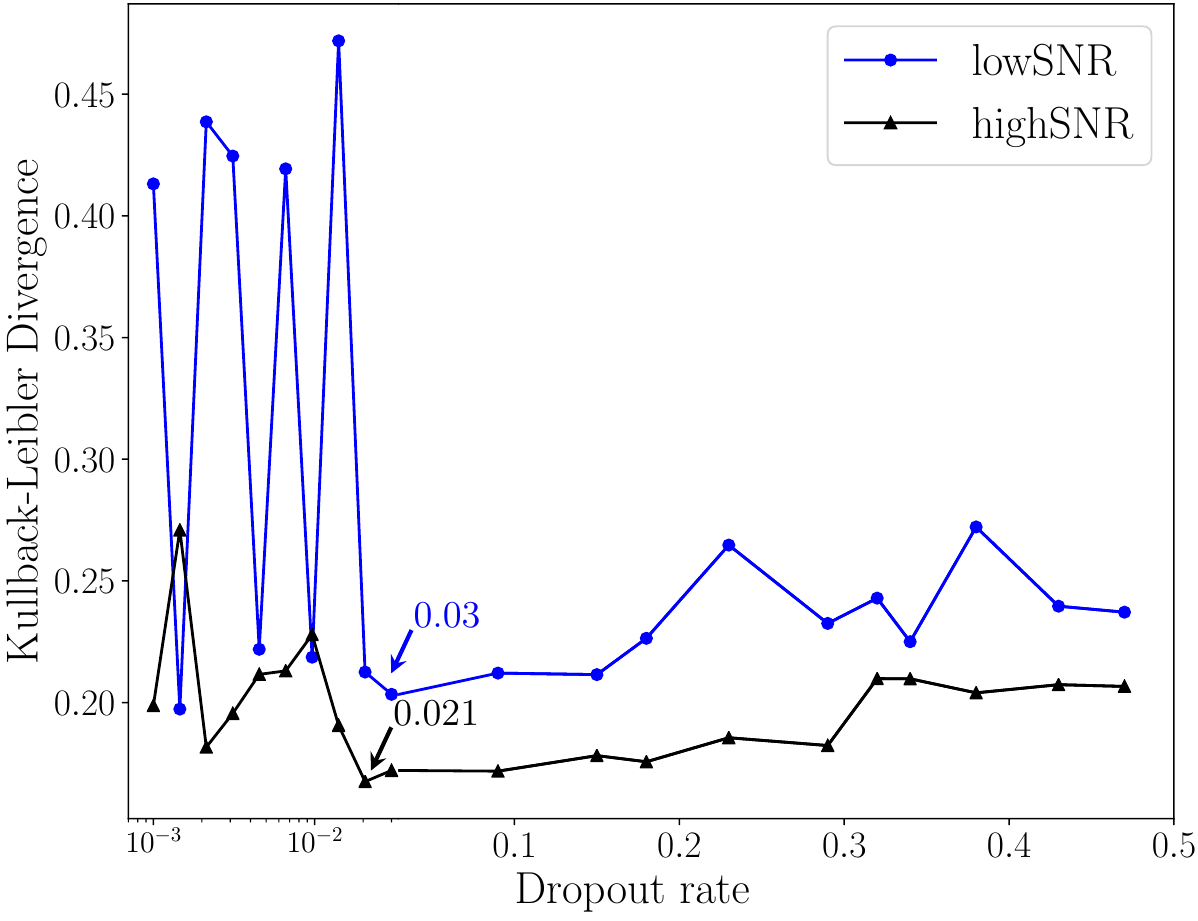}{0.50\textwidth}{\label{fig:KL divergence} (a) KL divergence measures the similarity of the distribution in one dimension, so we flatten the distribution of the three categories in soft labels into one dimension.}}
\gridline{\fig{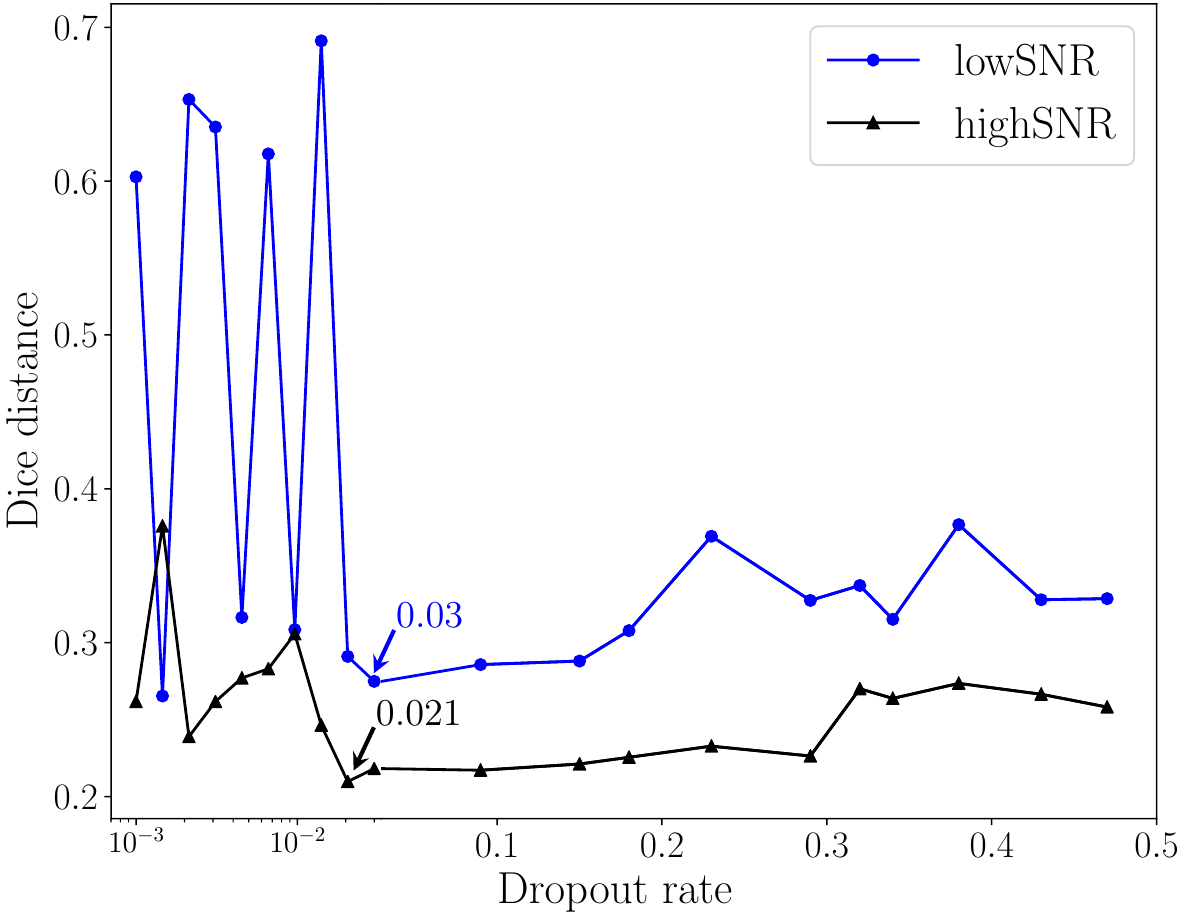}{0.50\textwidth}{\label{fig:Dice distance}(b) Dice distance can measure the similarity of the distribution in multiple dimensions, so the correspondence between the categories before and after inference can be preserved.}
          }
\caption{KL divergence and Dice distance for different dropout rates obtained by random search. The lower the scores of both metrics, the better the similarity of the predicted probability posterior distribution to the true soft-label probability distribution. Ten of the points are from the log space near dropout=0. We take the lowest point under stable performance as the good dropout rate. For the case of a low SNR (4$\sigma$), the optimal dropout rate is taken as 0.03; for the case of a high SNR (7$\sigma$), the optimal dropout rate can be taken between 0.021 and 0.3.}
\label{fig:KL and Dice}
\end{figure}

Figure~\ref{fig:KL and Dice} shows the KL divergence of our 20 random points with dropout rates between 0 and 0.5. These points include the 10 points found by random search and the 10 points taken out equivalently in their logarithmic space after localising to a smaller interval. Lower KL divergence and Dice distance values represent the more similar the two distributions are. We can see from the figure that the overfitting caused by the too-small dropout rate makes the scatter of the predicted posterior distribution with the ground truth distribution large, and the KL divergence value reaches a stable minimum as the dropout increases. Then, as the dropout rate increases, it leads to a lower capability of the model, which in turn leads to a predicted distribution that does not match the ground truth distribution. We find that the dropout can be taken to have the minimum scatter at a low SNR (4$\sigma$) equal to 0.03; while the optimal dropout value to minimize the scatter is 0.021 for high SNR (7$\sigma$), and the range of the optimal dropout can be extended to 0.021-0.3 for high SNR (7$\sigma$). The curves of the Dice distance also supported this suggestion for the dropout rate choice.
\newline

\begin{figure*}[!htb]
\centering
% \includesvg[width=1\textwidth]{figures/predict_probability_one_sample_of_each_in_whole_classes_highSNR.svg}
\includegraphics[width=1\textwidth]{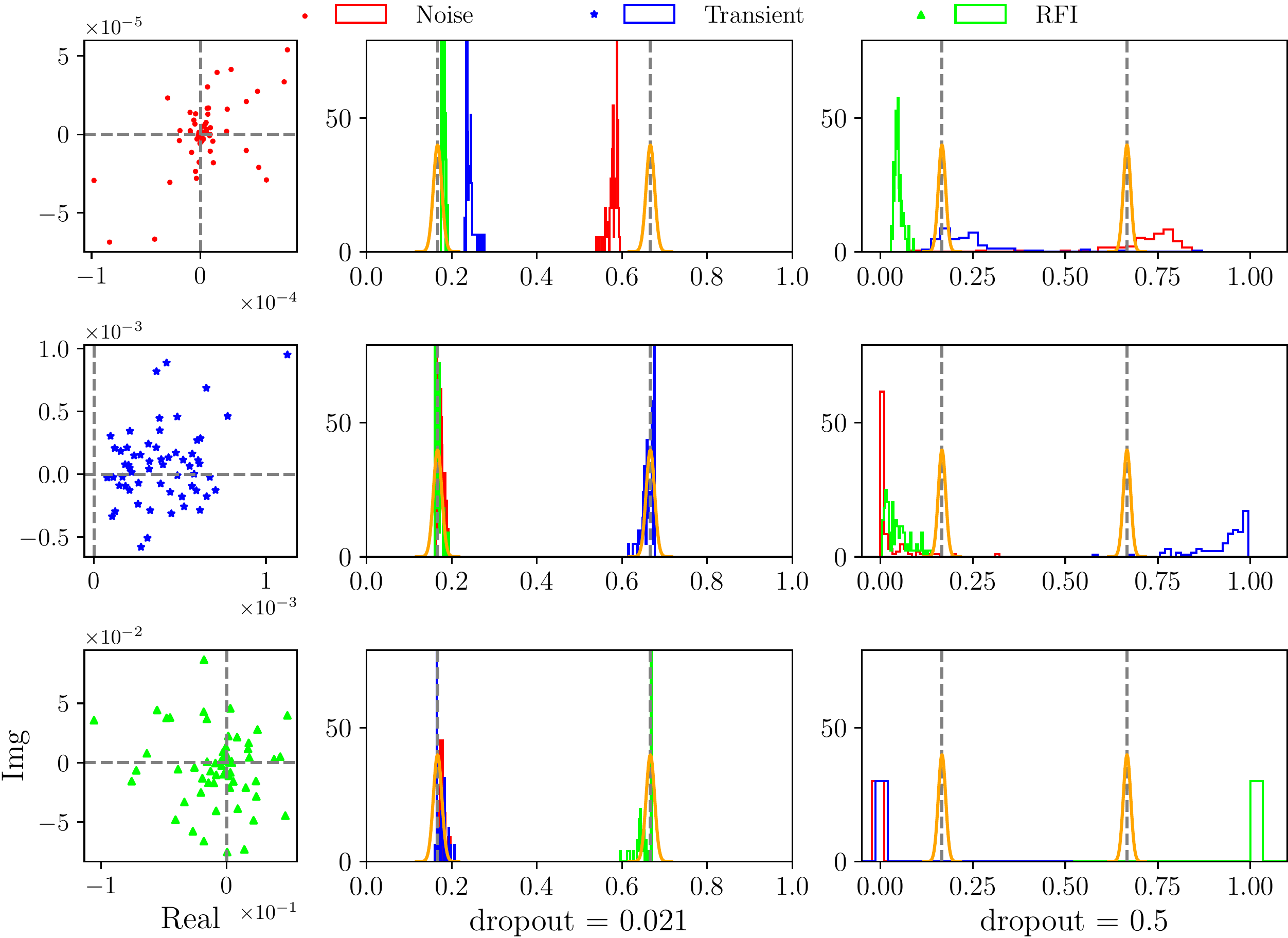}
\caption{\label{fig:sample highSNR}Prediction posteriors for noise, transient, RFI in high SNR (7$\sigma$). The left-hand column shows the random examples, with all its closure product points in the complex plane marked with red dots, blue stars and green triangles, respectively. The transient source posterior with the dropout rate set to 0.021 matches the ground truth distribution well both in terms of the center of the probability output and the confidence. The results are much worse when the dropout rate is set to 0.5. Its posterior distributions mismatch the ground truth distribution, which give either overconfident predictions or high uncertainty and more misclassifications. Here, two orange probability distribution curves have a variance of 0.01, centering at 2/3 and 1/6, respectively. The former represents the probability distribution of the primary categorical dimension in the ground truth soft labels of the test data, and the latter represents the probability distributions of the other two categorical dimensions.}
\end{figure*}

\begin{figure*}[!htb]
\centering
% \includesvg[width=1\textwidth]{figures/predict_probability_one_sample_of_each_in_whole_classes_lowSNR.svg}
\includegraphics[width=1\textwidth]{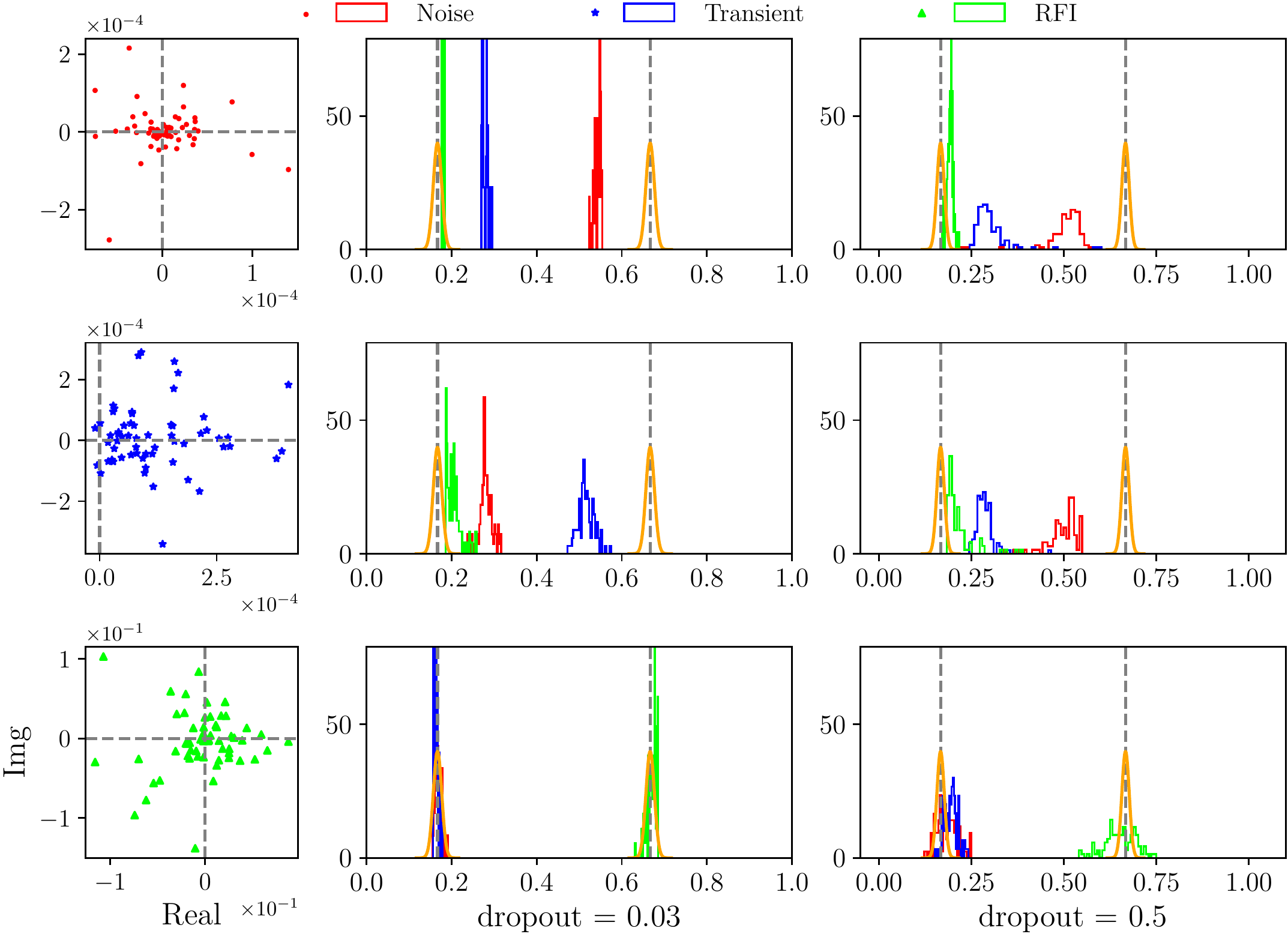}
\caption{\label{fig:sample lowSNR}As for Figure~\ref{fig:sample highSNR}, but showing posteriors for noise in low SNR level (4$\sigma$). The posteriors of transient source posterior with the dropout rate set to 0.03 starts to deviate slightly from the ground truth distribution in terms of both the center of the probability output and the confidence level. And the model gives even worse results when the dropout rate is set to 0.5. Compared to Figure~\ref{fig:sample highSNR}, the overconfidence prediction is reduced, but it still stays high prediction uncertainty and more false negatives.}
\end{figure*}

Next, in order to qualitatively provide examples of the posterior distribution matching, we randomly selected one example from each category of the noise, transient source, and RFI, and compared the posterior results for each category when we took the optimal dropout value as well as the dropout = 0.5. Figure~\ref{fig:sample highSNR} show the high SNR results, and Figure~\ref{fig:sample lowSNR} show the low SNR results. 
\newline  

As an example, in Figure~\ref{fig:sample highSNR}, each row is a randomly picked noise, transient, and RFI datum, respectively. The first column on the left shows the signal examples of closed products in the complex plane. The second column shows the predicted posterior distribution of soft labels after 100 forward passes when the dropout takes the optimal value of 0.021 (the horizontal axis indicates the probability values of the possible posteriors, and the vertical axis indicates the corresponding PDF). The fixed orange probability distribution curve (center value of 2/3 and variance of 0.01) is the distribution of the maximum pseudo-probability values in the real soft labels of the test data, i.e., the pseudo-probability values of the transient category locations; the other orange distribution curve with the center value of 1/6 and variance of 0.01 is the distribution of the corresponding noise and RFI location in the ground-truth soft labels of the test data. Blue is the probability posterior distribution of the category of transient source in the predicted soft labels; red is the probability posterior distribution of the noise category in the predicted soft labels; green is the probability posterior distribution of the RFI category in the predicted soft labels. The third column is similar to the second column, but it is the predicted results for dropout=0.5. Note that the Gaussian probability distribution of soft labels does not represent the physical properties of each signal component, and the prior information of such labels is only set to Gaussianity here for the convenience of the simulation. Ultimately, the degree to which the predicted PDFs of the soft labels match the ground truth PDFs is examined. 
\newline

By examining the independent examples, it can be seen that the centers of all posterior distributions are very close to or overlap with the ground truth distribution, when dropouts are taken to their optimal values in both high and low SNR cases, and the uncertainty of each distribution is also very similar to the ground truth distribution. Excellent matching is obtained for the inferences of transient source and RFI when dropout is taken as 0.021 in the high SNR case (7$\sigma$). It also holds for vast majority of their counterparts when dropout is taken as 0.03 in low SNR case (4$\sigma$), but some individual events become confused with noise. 
%The inference of the transient source shows such confusion (and more often than RFI, which cannot be showcased in figure yet worth to mention), and the posteriors also start to present imperfect matches in 
See Figure~\ref{fig:sample lowSNR}. This suggests that maximizing the SNR of the training data can improve classification accuracy. As for the noise inference, although it does not fit perfectly, the centers of the distributions of the noise category are also close to the center of the distribution of the ground truth soft labels, and its uncertainty is basically the same as the ground truth distribution. It is worth noting that the posterior values of the noise category for the two SNR levels discussed here (Figure~\ref{fig:sample highSNR} and \ref{fig:sample lowSNR}) do not perfectly reproduce the input ground truth, and the results in the low SNR case is worse than that in the high SNR case; however they are significantly better than the results for a drop out of 0.5.
This is additional evidence that high SNR data can improve the model's capability.
\newline 

Further, and more importantly, the models with the best dropout values for both high and low SNRs gave overall better predictions than the model with a standard dropout rate of 0.5. Dropout rates taking 0.5 typically show inconsistency with the ground truth distribution and more significant uncertainty in the posterior distribution. For the former bias, either can be seen in Figure~\ref{fig:sample highSNR}, when transient sources are classified with RFI at high SNR (7$\sigma$), which shows incredibly high confidence with deviation from the actual distribution (output classification probability values close to 1, coupled with small uncertainty), or can be found as more mis-classifications (transient is incorrectly classified as noise in Figure~\ref{fig:sample lowSNR}). With regard to the latter variance,  the greater predicted uncertainty can be mainly seen in both plots. It indicates that with the dropout rate set to 0.5, all these inferences behave far from the ground truth. This suggests that a uniform standard dropout rate value does not provide correct posterior distribution matching in a given scenario and with a given neural network structure, and that calibrating the dropout rate is a necessary step.

\subsection{Confidence Prediction on Semi-Real MWA Dataset}  
In our previous experiments, we built and optimized artificial intelligence networks to achieve high detection rates for test signals in a high discrimination feature space of closure product signals under simulated Gaussian noise and simulated transient source conditions, as well as to give predictions similar to ground truth labels. Due to the difficulty in obtaining large amounts of pertinent real labelled data, we can further test our trained models with more challenging data.

In this experiment, we consider observational data generated from a wide field-of-view transient source observation scenario using the MWA array. The dataset was constructed as a semi-real dataset containing real sky noise and a simulated transient model of single rising and falling flare. It is designed to mimic the recent discovery by \citet{MWA-transient}. In this demonstration, we set aside RFI. 
In the experiment presented here, we simulated 26 separate synthetic observation files each containing a transient source in the same sky noise background. The transient rose from zero to 8$\sigma$ over the sky noise on the 4th time step, then a further 4, then 2 and then 1$\sigma$ in the subsequent time steps. Then it rapidly fell by 8, then 4, 2, and 1$\sigma$, back to zero. This profile of inverse exponential rise and exponential decay we called the Shark-Fin.
Each data file contains a single transient source in different locations that were spread over the sky to reflect more realistic observational scenarios. At no time were these new MWA data included in the previous model training or validation experiments.
\newline

We tested this semi-real data with our best model, which had a dropout rate of 0.03 and was trained by the data with soft labels. Using the soft-labelled training data with slight uncertainty as augmented data will result in a more robust final model, and the inferred best confidence output made with this model would be close to the previously selected probability distributions for soft-label category elements, which is centered at 0.66 as well as the ones centered at 0.17.
\newline

In our experiments, we needed to manipulate the test MWA data prior to input to fit the requirements of the model's input dimension. The MWA data here are observed by 101 effective antennas with 512 frequency channels, which form 101 $\times$ 100/2 baseline pairs and 101 $\times$ 100 $\times$ 99/6 closure products. To best fit our fine-tuned model's input requirements and keep a high level of sensitivity of the signal, we made the following adjustments to the MWA test data. The 101 $\times$ 100 $\times$ 99/6 closure products were combined into 20 composite ones, binned to have an equal number in each bin, sorted on perimeter length. The 512 frequency channels were combined into three composite channels. However, if we had done this directly, the signal of the transient source far from the center of the observed phase of the MWA would have been severely diluted. The further away from the center, the more severe the signal loss. This is due to bandwidth smearing caused by the averaging of the frequency channels, and signals from longer baselines lose a significant amount of sensitivity. To reduce the loss of effective field of view due to the frequency averaging, we maximized the pre-averaging of the baseline channels by 16 to 32 composite frequency channels and then to form the composite closure products (each integrated closed product contains approximately 8332 original closure products). We expect the transients around the phase center to work well, but perhaps not so successfully for those further from the centre. In this demonstration, we hope to learn about the detection limit as a function of the positions of transient sources as well.
\newline

\begin{figure*}
\gridline{\fig{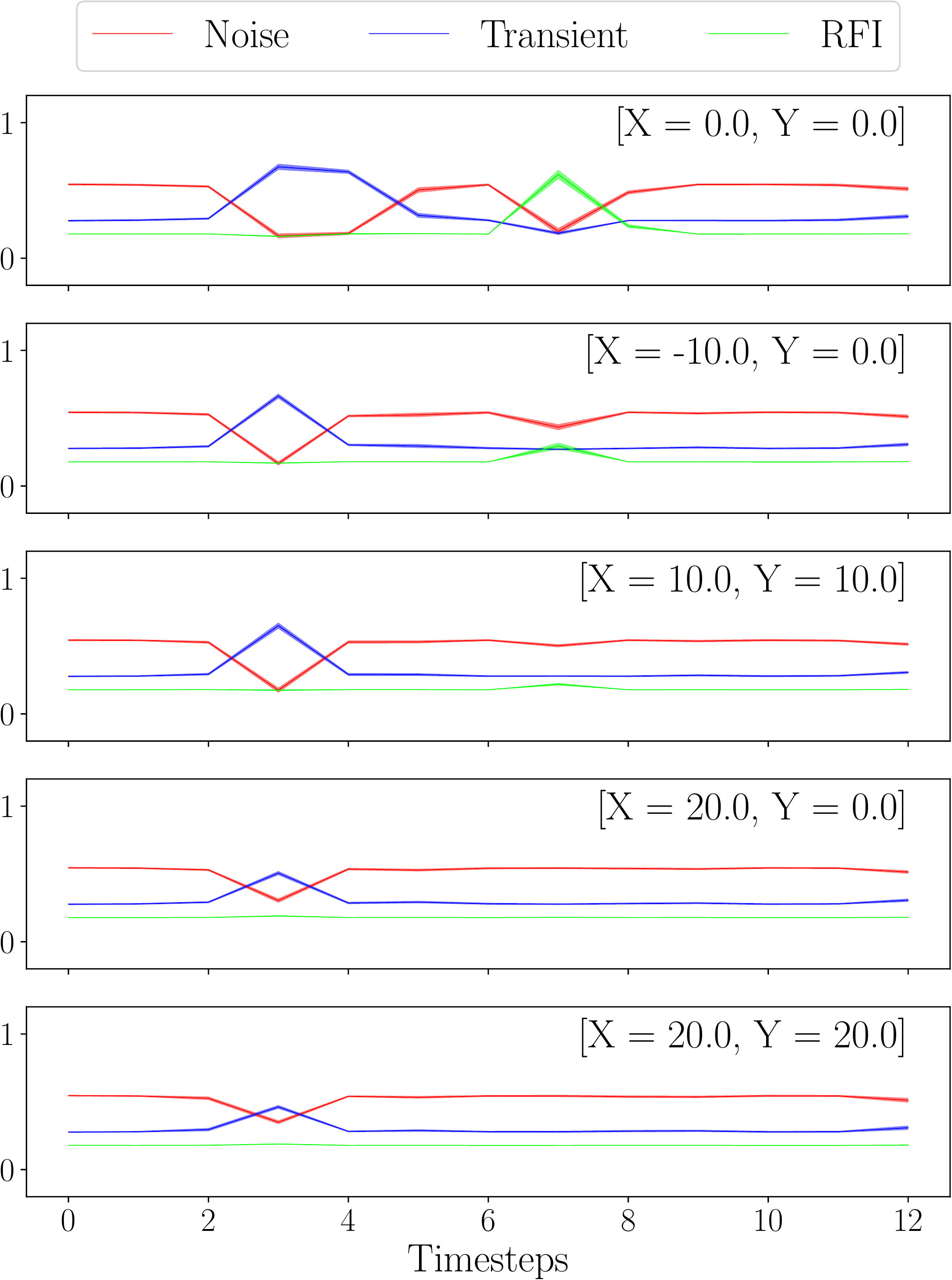}{0.46\textwidth}{\label{fig:0.03center}}
          \fig{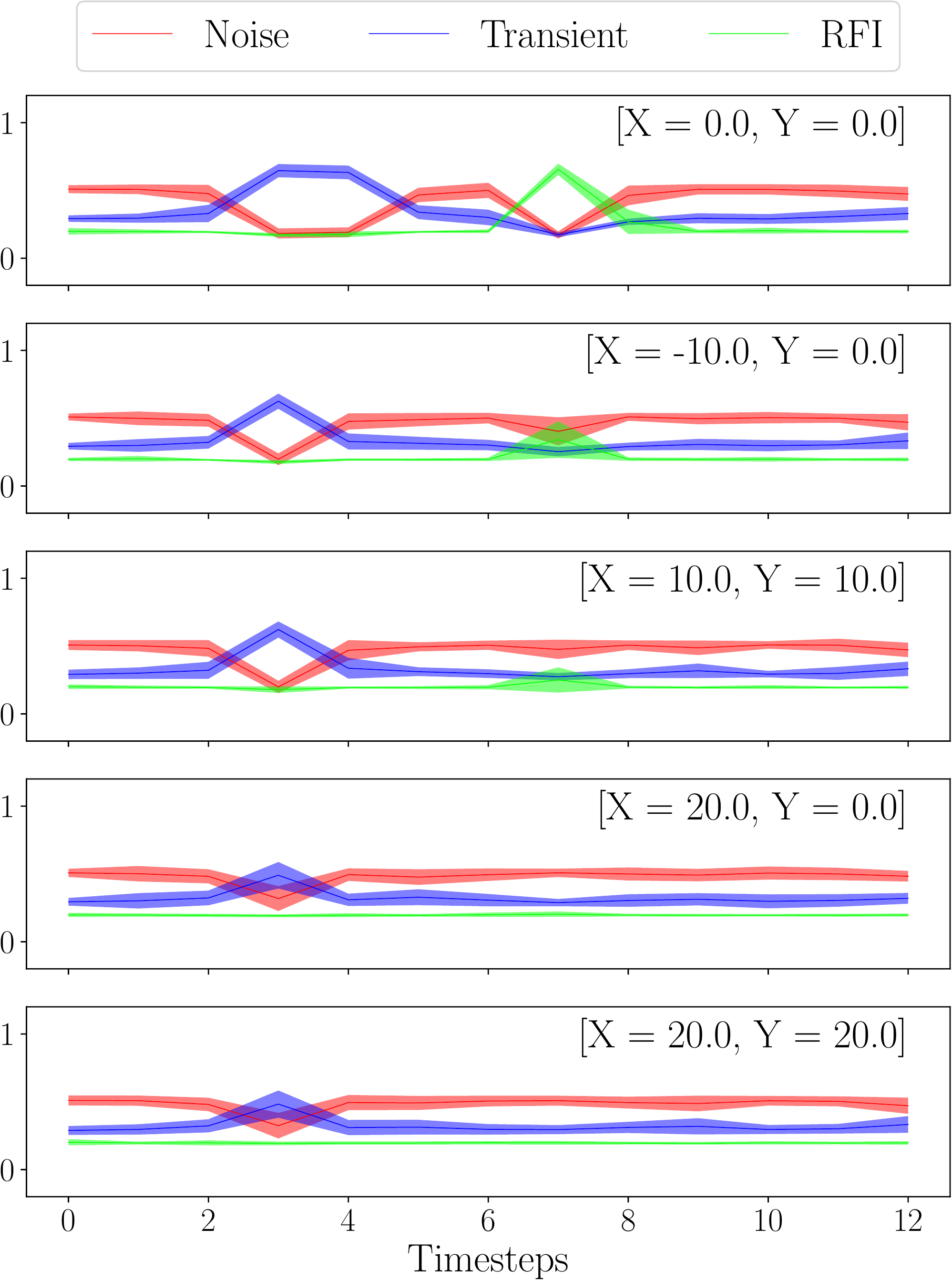}{0.46\textwidth}{\label{fig:0.03confusing}}
          }
\caption{The model predicts classification probabilities and uncertainty with optimal dropout rates of 0.03 (left) and 0.5 (right) for the entire 13-second observation segment on the MWA semi-real data set.
The upper right corner of each subplot is the offset of the angular distance between the transient event and the phase center of the observation. Here we choose to exhibit the positions whose predictions change dramatically. The highest value of predicted probability is 0.66 instead of 1 due to the central distribution of probability labels for each instance in the dataset used for model training with a value of 0.66.}
\label{fig:0.03transient probabilities}
\end{figure*}

Figure~\ref{fig:0.03transient probabilities} gives confidence curves for the evolution of the prediction categories over observation time for a part of 26 files containing transient sources. \textcolor{black}{For the 3rd differenced time step at all angular separations, the model gives a steep increase in the probability that this signal is a transient source, which means that the appearance of the transient source is accurately predicted. 
However the weaker transient signals are only detected for the smaller angular separation.}
During the decay of the transient source \textcolor{black}{the (7th differenced time step), 
the transient is misidentified as a RFI spike. This is because the simulation used for training had no labels for decaying signals; there was a transient or there was not.}
Comparing Figure~\ref{fig:0.03transient probabilities} (left) with Figure~\ref{fig:0.03transient probabilities} (right) which is for a dropout of 0.5, we see that it gives correct predictions for the appearance of transient sources close to the phase center, but the predictions have much greater uncertainties.
\newline

% \begin{figure*}
% \gridline{\fig{Plot_Datum_MWA_predicted_output_conphidence_uncertainty_dropout_0.03_part_1.pdf}{0.46\textwidth}{\label{fig:0.03center}Prediction in the phase center region.}
%           \fig{Plot_Datum_MWA_predicted_output_conphidence_uncertainty_dropout_0.03_part_complement_NoLegend.pdf}{0.46\textwidth}{\label{fig:0.03confusing}Prediction in the region where the probabilities of transient start to be confused with noise.}
%           }
% \caption{Predicted categorical probabilities and uncertainties on the MWA semi-real dataset for the model with the optimal dropout of 0.03.}
% \label{fig:0.03transient probabilities}
% \end{figure*}

% \begin{figure*}
% \gridline{\fig{Plot_Datum_MWA_predicted_output_conphidence_uncertainty_dropout_0.5_part_1.pdf}{0.46\textwidth}{\label{fig:0.5center}Prediction in the phase center region.}
%           \fig{Plot_Datum_MWA_predicted_output_conphidence_uncertainty_dropout_0.5_part_complement_NoLegend.pdf}{0.46\textwidth}{\label{fig:0.5confusing}Prediction in the region where the probabilities of transient start to be confused with noise.}
%           }
% \caption{Predicted categorical probabilities and uncertainties on the MWA semi-real dataset for the model with the conventional dropout of 0.5.}
% \label{fig:0.5transient probabilities}
% \end{figure*}

\begin{figure}[!htbp]
\centering
\includegraphics[width=0.49\textwidth]{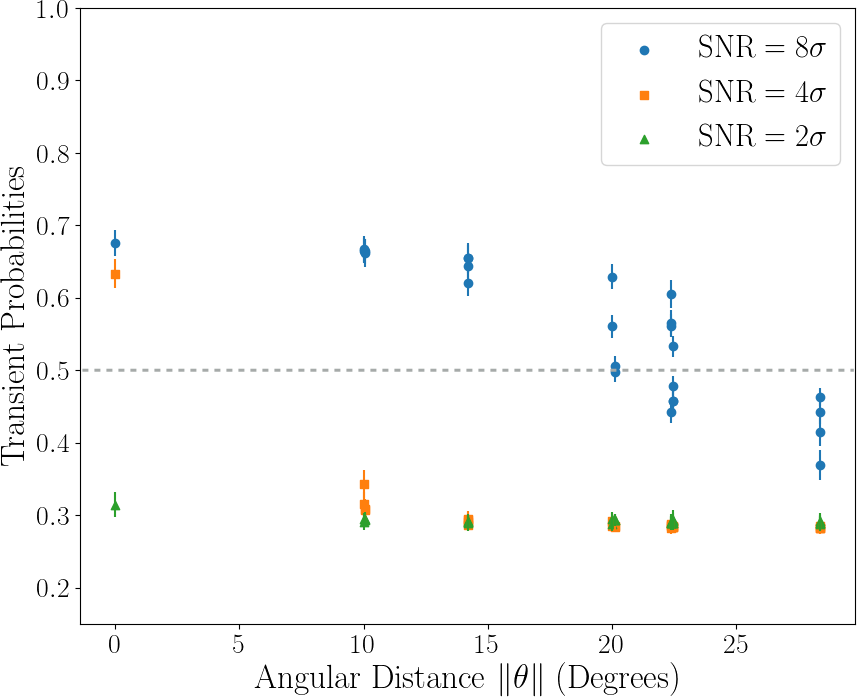}
\caption{\label{fig:AngularDistance_probabilities} Variation of the \textcolor{black}{inferred posterior probability of transient occurrence at the timestep four to six (corresponding to SNR of 8$\sigma$ in blue dots, 4$\sigma$ in orange squares, 2$\sigma$ in green triangles, respectively)} against the angular distance from the observational phase center. \textcolor{black}{The horizontal dashed line at 0.5 indicates that all the predictions of transients with probabilities above this threshold are identified as transients.}}
\end{figure}

Figure \ref{fig:AngularDistance_probabilities} gives the model prediction of the transient source as a function of angular separation and as a function of SNR. The model predicts the highest probability of the transient source when it appears at the observational phase center, for all SNR. For the strongest transient (8$\sigma$), the probability decreases as the transient source's location gradually moves away from the phase center, and the uncertainty gradually increases until the signals at the edge fail to be recognized as transient. For the mid-strength transient (4$\sigma$) the transient is detected close to the phase centre, but not at larger separations.
\newline

% Shark Fin SNR plot

% \begin{figure}[!htbp]
% \centering
% \includegraphics[width=0.45\textwidth,trim={7cm 6cm 7cm 1cm}]{figures/shark-fin_4paper.pdf}
% \caption{\label{fig:shark_fin_model} The light curve of the ``Shark fin" transient model. The differences between the two adjacent points are individual test samples. The SNRs of samples corresponding to the transient in the rising part are 8$\sigma$, 4$\sigma$, 2$\sigma$, 1$\sigma$, respectively. The SNRs of samples in the falling part are similar.}
% \end{figure}

% \begin{table}[h!]
% \centering
% \begin{tabular}{|l|l|l|}
% \hline
% SNR & Transient probability \\ \hline
% 1$\sigma$     & 0.306 $\pm$ 0.0494   \\ \hline
% 2$\sigma$    &  0.344 $\pm$ 0.0524  \\ \hline
% 4$\sigma$     &  0.632 $\pm$ 0.0565 \\ \hline
% 8$\sigma$    & 0.638 $\pm$ 0.0464  \\ \hline
% \end{tabular}
% \caption{Variation of the probability of transient occurrence against the input SNR at the observational phase center.}
% \label{table:transient_SNRvsProbability}
% \end{table}

We need to point out here that the current model does not accurately predict the falling portion of the transient signals even for the source in the phase center, but this is not to say that our solution is wrong or pointless. Instead, it is because we focused on the rising of transient sources during the initial training process, which is the only one that matters. Remember that in Equation~\ref{eqn:CP-Diff}, each timestep of closure product difference is generated by consecutive time instances $t_{k}$ and $t_{k-1}$. Only transient samples with positive features are labelled in our simulated training data and our current model can only detect the positive differences. In reality, as long as we can accurately detect the onset of transient sources, the fall is not that important. 
%Particularly considering that most of the falls are decay,  we can detect instant rises in a short time cadence under signal conditions such as the SNR of 7$\sigma$, but cannot detect slow falls that are not in one single cadence, which we can ignore completely. 
\newline

Therefore, we conclude that the optimal choice for one of the hyperparameters, the dropout rate, obtained by MC dropout approximating Bayesian optimization, allows the model to give small uncertainty and accurate inferences when performing transient source detection. The success of this experiment is significant for the practical construction of workflows. Rarely do we have a large number of similar yet labelled observations from different observation projects with different configurations and conditions for a neural network model to learn. In contrast, we can now build up a model by simulating close to real data. The resulting best model can perform efficient inference tasks in more severe scenarios, rather than requiring labelling to acquire large training sets of real observational data.
\newline

Given that the model can be changed and retrained for its input features, subsequent increases in the number of frequency channels will further reduce the effect of bandwidth smearing and allow for a larger effective field of view available for inference, GPU computing power permitting.

\section{Conclusion}
\label{sec:Conclusion}
In this paper, we investigated the feasibility of a machine learning approach to detect transients in correlated data. We proposed the idea of detecting transients by directly classifying three signal components: transients, RFI, and noise in correlated data with artificial intelligence neural networks. The closure products of the correlation function data corresponding to these three signal components have distinctly different statistical features. We built a simple yet efficient MLP (Multiple layer perceptron) with low network complexity to accomplish the recognition of transients. To explore the boundaries of the capabilities of the ML model, we simulated high SNR data sets (7$\sigma$) and low SNR data sets (4$\sigma$). We found that the F1-score for the classification is greater than 85\%, with the SNR at 7$\sigma$.
\newline

Also, to obtain the inferential uncertainty of the model in applying dropout rates, we proposed to apply a Monte Carlo dropout \citep{gal2016dropout} approximation Bayesian inference to estimate how well the predicted posterior matches the actual underlying probability distribution of the detected events at different dropout rates. We analyzed the confidence of the predictions after converting hard labelling output to soft labelling output with pseudo-probability values. By comparing the prediction results with the commonly adopted dropout value fixed at 0.5, we found that the model with the best dropout rate predicts posteriors that better match the pre-defined ground truth distribution.
\newline

We tested the data stream with the proposed well-trained model with the best dropout value on more challenging semi-real data (simulated transient source and real noise). The results show that our model detected the occurrence of the transient source with a high probability value, and the uncertainty was significantly reduced compared to the results given by the popular choice of 0.5. The success of the model’s prediction on semi-real data shows that we do not have to struggle with finding a large number of correctly labeled real observations as training data in future, but can train a good practical model using reasonable simulated data.
\textcolor{black}{%However, as our model did not include labelled falling transients, it was unable to detect such events. Training new models that included these labels would allow transient variability as well as occurrence to be identified.
However, we did not include the feature of the decay of transient sources with negative values of closure products into the training data. In future work, we can add this feature forming the entire life of the transient source to the training data for the model learning.}
\newline

In addition, we note that in the construction of the training data feature vectors,
in order to adapt to the arithmetic power of the GPU and to improve the sensitivity of the signals of the input vectors, we merged and grouped the information of the closure products along the frequency channels. Such manipulation distorts the flux of transient sources located far from the observational phase center, resulting in inaccurate and unstable predictions for these edged sources. 
At this point, our trained model is able to detect the rising phase transient source well. With a SNR of eight or above we confidently detect the transients out to a radius of 20$\deg$.
For weaker sources the angular extent to which we are sensitive is significantly reduced to about a radius of 5$\deg{}$ at 4$\sigma$.
Our experiments also help us to find the range of inferable regions, which we can overcome by rotating the observational phase center and performing a parallel patch detection in future studies.
\newline

In the work of this paper, we treat transient source detection as a classification problem. In subsequent work, we will model transient source detection as a multi-dimensional semantic segmentation problem, which will better fit the scenario of capturing transient sources within the time-domain observation data stream. The successful exploration of the classification problem will significantly increase the feasibility of subsequent semantic segmentation.

\section*{acknowledgments}
Xia Zhang acknowledges financial support from the Chinese Scholarship Council (CSC).


\begin{thebibliography}{}
\expandafter\ifx\csname natexlab\endcsname\relax\def\natexlab#1{#1}\fi
\providecommand{\url}[1]{\href{#1}{#1}}
\providecommand{\dodoi}[1]{doi:~\href{http://doi.org/#1}{\nolinkurl{#1}}}
\providecommand{\doeprint}[1]{\href{http://ascl.net/#1}{\nolinkurl{http://ascl.net/#1}}}
\providecommand{\doarXiv}[1]{\href{https://arxiv.org/abs/#1}{\nolinkurl{https://arxiv.org/abs/#1}}}

\bibitem[{Akeret {et~al.}(2017)Akeret, Chang, Lucchi, \&
  Refregier}]{akeret2017radio}
Akeret, J., Chang, C., Lucchi, A., \& Refregier, A. 2017, Astronomy and
  computing, 18, 35

\bibitem[{Baron(2019)}]{baron2019machine}
Baron, D. 2019, arXiv preprint arXiv:1904.07248

\bibitem[{Bergstra \& Bengio(2012)}]{bergstra2012random}
Bergstra, J., \& Bengio, Y. 2012, Journal of machine learning research, 13

\bibitem[{Bhat(2011)}]{bhat2011searches}
Bhat, N. 2011, arXiv preprint arXiv:1110.2562

\bibitem[{Bramich(2008)}]{bramich2008new}
Bramich, D. 2008, Monthly Notices of the Royal Astronomical Society: Letters,
  386, L77

\bibitem[{Burd {et~al.}(2018)Burd, Mannheim, M{\"a}rz, Ringholz, Kappes, \&
  Kadler}]{burd2018detecting}
Burd, P.~R., Mannheim, K., M{\"a}rz, T., {et~al.} 2018, Astronomische
  Nachrichten, 339, 358

\bibitem[{Connor \& van Leeuwen(2018)}]{connor2018applying}
Connor, L., \& van Leeuwen, J. 2018, The Astronomical Journal, 156, 256

\bibitem[{Cornwell(1987)}]{cornwell1987radio}
Cornwell, T. 1987, Astronomy and Astrophysics, 180, 269

\bibitem[{Cornwell {et~al.}(1999)Cornwell, Braun, \&
  Briggs}]{cornwell1999deconvolution}
Cornwell, T., Braun, R., \& Briggs, D.~S. 1999, in Synthesis Imaging in Radio
  Astronomy II, Vol. 180, 151

\bibitem[{{CSIRO-ATNF}(2020)}]{SKA2020}
{CSIRO-ATNF}. 2020, {The Square Kilometre Array},
  \url{https://www.atnf.csiro.au/projects/ska/index.html}

\bibitem[{Czech {et~al.}(2018)Czech, Mishra, \& Inggs}]{czech2018cnn}
Czech, D., Mishra, A., \& Inggs, M. 2018, Astronomy and computing, 25, 52

\bibitem[{{Dewdney}(2013)}]{ska-parameter}
{Dewdney}, P.~E. 2013, { SKA-TEL-SKO-DD-001-1\_BaselineDesign.pdf},
  \dodoi{10.1109/JPROC.2009.2021005}

\bibitem[{Dice(1945)}]{dice1945measures}
Dice, L.~R. 1945, Ecology, 26, 297

\bibitem[{Dodson {et~al.}(2014)Dodson, Kim, Sohn, Rioja, Jung, Seymour, \&
  Raja}]{dodson2014kava}
Dodson, R., Kim, C., Sohn, B., {et~al.} 2014, Publications of the Astronomical
  Society of Japan, 66

\bibitem[{Gal \& Ghahramani(2016)}]{gal2016dropout}
Gal, Y., \& Ghahramani, Z. 2016, in international conference on machine
  learning, PMLR, 1050--1059

\bibitem[{Grandini {et~al.}(2020)Grandini, Bagli, \&
  Visani}]{grandini2020metrics}
Grandini, M., Bagli, E., \& Visani, G. 2020, arXiv-preprint-arXiv:2008.05756

\bibitem[{H{\"o}gbom(1974)}]{hogbom1974aperture}
H{\"o}gbom, J. 1974, Astronomy and Astrophysics Supplement Series, 15, 417

\bibitem[{{Hurley-Walker} {et~al.}(2022){Hurley-Walker}, {Zhang}, {Bahramian},
  {McSweeney}, {O'Doherty}, {Hancock}, {Morgan}, {Anderson}, {Heald}, \&
  {Galvin}}]{MWA-transient}
{Hurley-Walker}, N., {Zhang}, X., {Bahramian}, A., {et~al.} 2022, \nat, 601,
  526, \dodoi{10.1038/s41586-021-04272-x}

\bibitem[{Jaynes(2003)}]{jaynes2003probability}
Jaynes, E.~T. 2003, Probability theory: The logic of science (Cambridge
  university press)

\bibitem[{Kingma \& Ba(2014)}]{kingma2014adam}
Kingma, D.~P., \& Ba, J. 2014, arXiv preprint arXiv:1412.6980

\bibitem[{Kulkarni(1989)}]{kulkarni1989self}
Kulkarni, S.~R. 1989, Astronomical Journal, 98, 1112

\bibitem[{Law \& Bower(2012)}]{law2012all}
Law, C.~J., \& Bower, G.~C. 2012, The Astrophysical Journal, 749, 143

\bibitem[{Michilli {et~al.}(2018)Michilli, Hessels, Lyon, Tan, Bassa, Cooper,
  Kondratiev, Sanidas, Stappers, \& van Leeuwen}]{michilli2018single}
Michilli, D., Hessels, J., Lyon, R., {et~al.} 2018, Monthly Notices of the
  Royal Astronomical Society, 480, 3457

\bibitem[{Milletari {et~al.}(2016)Milletari, Navab, \& Ahmadi}]{milletari2016v}
Milletari, F., Navab, N., \& Ahmadi, S.-A. 2016, in 2016 fourth international
  conference on 3D vision (3DV), IEEE, 565--571

\bibitem[{Murphy {et~al.}(2013)Murphy, Chatterjee, Kaplan, Banyer, Bell,
  Bignall, Bower, Cameron, Coward, Cordes, {et~al.}}]{murphy2013vast}
Murphy, T., Chatterjee, S., Kaplan, D.~L., {et~al.} 2013, Publications of the
  Astronomical Society of Australia, 30

\bibitem[{Offringa {et~al.}(2014)Offringa, McKinley, Hurley-Walker, Briggs,
  Wayth, Kaplan, Bell, Feng, Neben, Hughes, {et~al.}}]{offringa2014wsclean}
Offringa, A., McKinley, B., Hurley-Walker, N., {et~al.} 2014, Monthly Notices
  of the Royal Astronomical Society, 444, 606

\bibitem[{Riess {et~al.}(1998)Riess, Filippenko, Challis, Clocchiatti, Diercks,
  Garnavich, Gilliland, Hogan, Jha, Kirshner,
  {et~al.}}]{riess1998observational}
Riess, A.~G., Filippenko, A.~V., Challis, P., {et~al.} 1998, The Astronomical
  Journal, 116, 1009

\bibitem[{Rogers {et~al.}(1995)Rogers, Doeleman, \& Moran}]{rogers1995fringe}
Rogers, A.~E., Doeleman, S.~S., \& Moran, J.~M. 1995, The Astronomical Journal,
  109, 1391

\bibitem[{Rowlinson {et~al.}(2019)Rowlinson, Stewart, Broderick, Swinbank,
  Wijers, Carbone, Cendes, Fender, van~der Horst, Molenaar,
  {et~al.}}]{rowlinson2019identifying}
Rowlinson, A., Stewart, A.~J., Broderick, J.~W., {et~al.} 2019, Astronomy and
  Computing, 27, 111

\bibitem[{Stewart {et~al.}(2016)Stewart, Fender, Broderick, Hassall,
  Mu{\~n}oz-Darias, Rowlinson, Swinbank, Staley, Molenaar, Scheers,
  {et~al.}}]{stewart2016lofar}
Stewart, A., Fender, R., Broderick, J.~W., {et~al.} 2016, Monthly Notices of
  the Royal Astronomical Society, 456, 2321

\bibitem[{Sudre {et~al.}(2017)Sudre, Li, Vercauteren, Ourselin, \&
  Cardoso}]{sudre2017generalised}
Sudre, C.~H., Li, W., Vercauteren, T., Ourselin, S., \& Cardoso, M.~J. 2017, in
  Deep learning in medical image analysis and multimodal learning for clinical
  decision support (Springer), 240--248

\bibitem[{Trott {et~al.}(2011)Trott, Wayth, Macquart, \&
  Tingay}]{trott2011source}
Trott, C.~M., Wayth, R.~B., Macquart, J.-P.~R., \& Tingay, S.~J. 2011, The
  Astrophysical Journal, 731, 81

\bibitem[{Walmsley {et~al.}(2020)Walmsley, Smith, Lintott, Gal, Bamford,
  Dickinson, Fortson, Kruk, Masters, Scarlata, {et~al.}}]{walmsley2020galaxy}
Walmsley, M., Smith, L., Lintott, C., {et~al.} 2020, Monthly Notices of the
  Royal Astronomical Society, 491, 1554

\bibitem[{Wilson {et~al.}(2013)Wilson, Rohlfs, \&
  H{\"u}ttemeister}]{wilson2013interferometers}
Wilson, T.~L., Rohlfs, K., \& H{\"u}ttemeister, S. 2013, in Tools of Radio
  Astronomy (Springer), 237--288

\bibitem[{Wu {et~al.}(2019)Wu, Wong, Rudnick, Shabala, Alger, Banfield, Ong,
  White, Garon, Norris, {et~al.}}]{wu2019radio}
Wu, C., Wong, O.~I., Rudnick, L., {et~al.} 2019, Monthly Notices of the Royal
  Astronomical Society, 482, 1211

\bibitem[{Zackay {et~al.}(2016)Zackay, Ofek, \& Gal-Yam}]{zackay2016proper}
Zackay, B., Ofek, E.~O., \& Gal-Yam, A. 2016, The Astrophysical Journal, 830,
  27

\bibitem[{Zhang {et~al.}(2018)Zhang, Gajjar, Foster, Siemion, Cordes, Law, \&
  Wang}]{zhang2018fast}
Zhang, Y.~G., Gajjar, V., Foster, G., {et~al.} 2018, The Astrophysical Journal,
  866, 149

\end{thebibliography}
\end{document}